\documentclass[12pt,a4paper,english]{article}
\pdfoutput=1

\usepackage{epsf}
\usepackage{amsmath,amsfonts,amsthm,amssymb,mathtools}
\usepackage{graphicx}
\usepackage{color}
\usepackage{psfrag}
\usepackage{cite}

\usepackage{caption}
\usepackage{subcaption}

\usepackage{lmodern}
\usepackage[english]{babel}

\usepackage[bookmarks,bookmarksnumbered]{hyperref}
\hypersetup{colorlinks,allcolors=blue}
\hypersetup{pdftitle={M-theory interpretation of the real topological string},pdfauthor={N. Piazzalunga, A. M. Uranga}}
\hypersetup{linktocpage=true}

\usepackage{cleveref}
\usepackage{tikz}
\usetikzlibrary{decorations.markings,arrows,backgrounds}
\usepackage{enumitem}
\setlist[enumerate,1]{label=(\roman*)}

\usepackage{ifpdf}

\usepackage[width=15.5cm,height=23.5cm,centering]{geometry}

\newcommand{\bmat}{\left(\begin{array}}
\newcommand{\emat}{\end{array}\right)}

\def\yzero{\smash{\hbox{$y\kern-4pt\raise1pt\hbox{${}^\circ$}$}}}

\renewcommand{\a}{\alpha}
\renewcommand{\b}{\beta}
\newcommand{\ad}{\dot\alpha}
\newcommand{\bd}{\dot\beta}
\newcommand{\vep}{\varepsilon}

\def\beq{\begin{equation}}
\def\eeq{\end{equation}}
\def\beqa{\begin{eqnarray}}
\def\eeqa{\end{eqnarray}}

\def\-{\hphantom{-}}
\def\s2{\frac{1}{\sqrt2}}

\def\IF{\relax{\rm I\kern-.18em F}}
\def\II{\relax{\rm I\kern-.18em I}}

\def\FF{\mathcal{F}}

\def\Dsl{\,\raise.15ex\hbox{/}\mkern-13.5mu D} 

\def\NN{{\cal N}}
\def\IC{{\bf C}}
\def\IS{{\bf S}}
\def\IR{{\bf R}}
\def\IZ{{\bf Z}}
\def\IX{{\bf X}}
\def\IT{{\bf T}}

\def\CP{{\bf CP}}
\def\RP{{\bf RP}}

\def\R{{\cal R}}

\newcommand{\op}{\mathcal{O}}
\renewcommand{\R}{\mathrm{R}}
\renewcommand{\L}{\mathrm{L}}
\newcommand{\SU}{\mathrm{SU}}
\newcommand{\SO}{\mathrm{SO}}
\newcommand{\U}{\mathrm{U}}
\newcommand{\ud}{\mathrm{d}}
\newcommand{\hil}{\mathcal{H}}
\newcommand{\im}{\begin{pmatrix}}
\newcommand{\fm}{\end{pmatrix}}
\newcommand{\bz}{\overline{z}}
\DeclareMathOperator{\tr}{tr}

\newcommand{\ov}[1]{\overline{#1}}
\newcommand{\verteq}{\rotatebox{90}{$\,=$}}
\newcommand{\equalto}[2]{\underset{\scriptstyle\overset{\mkern4mu\verteq}{#2}}{#1}}
\newcommand\iu{\mathrm{i}}
\newcommand{\ms}{\mathcal{M}}
\DeclareMathOperator{\jac}{Jac}
\newcommand{\hg}{{\hat{g}}}
\def\hs{\hat{\Sigma}}

\newcommand{\tgw}{\tilde{\mathrm{GW}}}
\newcommand{\tgv}{\tilde{\mathrm{GV}}}
\newcommand{\gv}{\mathrm{GV}}
\newcommand{\hgvs}{\hat{\mathrm{GV}}}
\newcommand{\hgv}{\hat{\mathrm{GV}}_{\hat g,\beta}}
\newcommand{\gvp}{\mathrm{GV}'^+_{\hat g,\beta}}
\newcommand{\gvm}{\mathrm{GV}'^-_{\hat g,\beta}}

\theoremstyle{plain}

\newtheorem{thm}{Theorem}[section]

\catcode`\@=11   
\newdimen\@rotdimen
\newbox\@rotbox  

\def\@vspec#1{\special{ps:#1}}
\def\@rotstart#1{\@vspec{gsave currentpoint currentpoint translate
   #1 neg exch neg exch translate}}
\def\@rotfinish{\@vspec{currentpoint grestore moveto}}
\def\@rotr#1{\@rotdimen=\ht#1\advance\@rotdimen by\dp#1%
   \hbox to\@rotdimen{\hskip\ht#1\vbox to\wd#1{\@rotstart{90 rotate}%
   \box#1\vss}\hss}\@rotfinish}
\def\@rotl#1{\@rotdimen=\ht#1\advance\@rotdimen by\dp#1%
   \hbox to\@rotdimen{\vbox to\wd#1{\vskip\wd#1\@rotstart{270 rotate}%
   \box#1\vss}\hss}\@rotfinish}%
\def\@rotu#1{\@rotdimen=\ht#1\advance\@rotdimen by\dp#1%
   \hbox to\wd#1{\hskip\wd#1\vbox to\@rotdimen{\vskip\@rotdimen
   \@rotstart{-1 dup scale}\box#1\vss}\hss}\@rotfinish}%
\def\@rotf#1{\hbox to\wd#1{\hskip\wd#1\@rotstart{-1 1 scale}%
   \box#1\hss}\@rotfinish}%
\def\rotate{\@ifnextchar[{\@rotate}{\@rotate[l]}}
\def\@rotate[#1]#2{\setbox\@rotbox=\hbox{#2}\@nameuse{@rot#1}\@rotbox}

\catcode`\@=12

\begin{document}


\numberwithin{equation}{section}

\begin{titlepage}
\thispagestyle{empty}

\rightline{IFT-UAM/CSIC-14-044}
\rightline{SISSA 30/2014/MATE}
\vspace{3mm}

\newcommand{\email}[1]{\href{mailto:#1}{\nolinkurl{#1}}}

\def\aone{\sharp}
\def\atwo{\clubsuit}
\def\athree{\spadesuit}

\makeatletter
\newcommand*{\wackyfn}[1]{\expandafter\@wackyfn\csname c@#1\endcsname}
\newcommand*{\@wackyfn}[1]{\ifcase#1\or$\aone$\or$\atwo$\or$\athree$\else\@ctrerr\fi}
\renewcommand\thefootnote{\wackyfn{footnote}}
\makeatother

\begin{center}
\LARGE{\bf M-theory interpretation\\of the real topological string\\[12mm]}
\large{Nicol\`o Piazzalunga%
\footnote{Email: \email{nicolo.piazzalunga@sissa.it}}
\qquad
Angel M.~Uranga%
\footnote{Email: \email{angel.uranga@uam.es}}
\\[4mm]}
\small{\it $^\aone$International School for Advanced Studies (SISSA)\\[-0.3em]
via Bonomea 265, 34136 Trieste, ITALY\\[1.5mm]
$^\aone$Istituto Nazionale di Fisica Nucleare (INFN), Sezione di Trieste\\[1.5mm]
$^\atwo$Instituto de F\'{\i}sica Te\'orica IFT-UAM/CSIC\\[-0.3em]
C/ Nicol\'as Cabrera 13-15, Universidad Aut\'onoma de Madrid, 28049 Madrid, SPAIN}\\[15mm]
\end{center}

\begin{abstract}
\normalsize
We describe the type IIA physical realization of the unoriented topological string introduced by Walcher,
describe its M-theory lift,
and show that it allows to compute
the open and unoriented topological amplitude in terms of one-loop diagram of BPS M2-brane states.
This confirms and allows to generalize the conjectured BPS integer expansion of the topological amplitude.
The M-theory lift of the orientifold is freely acting on the M-theory circle,
so that integer multiplicities are a weighted version
of the (equivariant subsector of the) original closed oriented Gopakumar-Vafa invariants.
The M-theory lift also provides new perspective on the topological tadpole cancellation conditions.
We finally comment on the M-theory version of other unoriented topological strings,
and  clarify certain misidentifications in earlier discussions in the literature.

\end{abstract}
\end{titlepage}

\newpage
\hypersetup{pageanchor=true}

\setcounter{page}{2}
\pagestyle{plain}
\renewcommand{\thefootnote}{\arabic{footnote}}
\setcounter{footnote}{0}


\tableofcontents

\section{Introduction}

Topological string theory is a fertile arena of interplay between physics and mathematics.
A prominent example is the physics-motivated reformulation of the topological A-model on a threefold $\IX_6$
in terms of integer multiplicities of BPS states in the 5d compactification of M-theory on $\IX_6$
\cite{Gopakumar:1998ii,Gopakumar:1998jq},
and the corresponding mathematical reformulation of the (in general fractional) Gromov-Witten invariants
in terms of the {\em integer} Gopakumar-Vafa invariants
(see also \cite{ionel}).

A natural generalization is to consider A-models with different worldsheet topologies. In particular, there is a similar story for the open topological A-model,
in which worldsheets are allowed to have boundaries mapped to a lagrangian 3-cycle in $\IX_6$,
and which via lift to M-theory admits an open BPS invariant expansion \cite{Ooguri:1999bv}.
There has also been substantial work to define unoriented topological A-models,
for instance in terms of the so-called real topological strings \cite{Walcher:2007qp,Krefl:2009mw,Krefl:2009md,Krefl:2008sj}.
The latter was proposed to require a specific open string sector for consistency,
and conjectured to admit a BPS-like expansion ansatz, although no physical derivation in terms of M-theory was provided.
Conversely, although some unoriented topological models have been proposed directly from the M-theory picture \cite{Sinha:2000ap,Acharya:2002ag,Bouchard:2004ri,Bouchard:2004iu,Aganagic:2012au}, they do not correspond to this real topological string.

In this paper we fill this gap, construct the physical theory corresponding to the real topological string,
and show that its M-theory lift reproduces the topological string partition function in terms of certain BPS invariants,
which we define and show to be the equivariant subsector of the corresponding closed oriented Gopakumar-Vafa invariants.
Along the way, the M-theory picture sheds new light into certain peculiar properties of the topological model,
like the so-called tadpole cancellation condition,
which requires combining open and unoriented worldsheets in order to produce well-defined amplitudes and integer invariants.
Although \cite{Walcher:2007qp} focused on the quintic and other simple examples
(see also \cite{Krefl:2009mw,Krefl:2009md,Krefl:2008sj}),
we keep the discussion general, using these examples only for illustration at concrete points.

The paper is organized as follows.
In \cref{sec:gv} we review the Gopakumar-Vafa reformulation of the closed oriented topological A-model.
In \cref{unoriented-gv} we review the properties of the real topological string.
In \cref{sec:mth} we present the physical IIA theory corresponding to this topological model
and construct its M-theory lift (\cref{sec:lift}),
compute the partition function in terms of M-theory BPS invariants (\cref{sec:bps}),
and describe the M-theory explanation of the tadpole cancellation conditions (\cref{sec:tadpole}).
In \cref{sec:extensions} we describe related systems,
by the inclusion of additional brane pairs (\cref{sec:extra-pairs}),
or by using other orientifold plane structures (\cref{sec:others});
in this respect, we clarify certain misidentifications of the M-theory lifts in the earlier literature on unoriented topological models.
Finally, \cref{sec:conclu} contains our conclusions.
\Cref{sec:walcher} reviews the basics of the real topological string,
while \cref{sec:couplings} discusses the physical couplings computed by the real topological string.

\section{Review of Gopakumar-Vafa expansion}
\label{sec:gv}

We start with a brief review of the Gopakumar-Vafa interpretation of the closed oriented topological string
in terms of BPS states in M-theory \cite{Gopakumar:1998ii,Gopakumar:1998jq}.

The 4d compactification of type IIA on a CY threefold $\IX_6$ provides a physical realization of the topological A-model on $\IX_6$,
whose genus $g$ partition function $F_g(t_i)$,
which depends on the K\"ahler moduli $t_i$, computes the F-term
\beq
\int \ud^4 x \int \ud^4\theta \, F_g(t_i) \left(\mathcal W^2\right)^g \to
\int \ud^4 x \, F_g(t_i) F_+^{2g-2} R_+^2
\label{R2}
\eeq
(where the second expression applies for $g>1$ only).
Here we have used the $\NN=2$ Weyl multiplet, schematically ${\cal {W}}=F_+ + \theta^2 R_+ + \cdots$,
with $F_+$, $R_+$ being the self-dual components of the graviphoton and curvature 2-form, respectively.
These contributions are summed up if we turn on a self-dual graviphoton background in the four non-compact dimensions
\beq
F_+ = \frac\lambda2 \ud x^1 \wedge \ud x^2 +\frac\lambda2 \ud x^3 \wedge \ud x^4.
\eeq
The sum is given by the total A-model partition function, with coupling $\lambda$
\beq
\FF(t_i) = \sum_{g=0}^{\infty} \lambda^{2g-2} F_g(t_i).
\eeq
There is an alternative way to compute this same quantity,
by considering the lift of the IIA configuration to M-theory, as follows.
We start with the 5d compactification of M-theory on $\IX_6$.
There is a set of massive half BPS particle states,
given by either the dimensional reduction of 11d graviton multiplets,
or by M2-branes wrapped on holomorphic 2-cycles. These states are characterized by their quantum numbers under the 5d little group $\SU(2)_\L\times \SU(2)_\R$.
Note that at the classical level, each such particle can have a classical moduli space,
but at the quantum level there is only a discrete set of ground states,
which provide the BPS particle states we are interested in.
For instance, an 11d particle (such as the 11d graviton) has a classical moduli space given by $\IX_6$ itself,
but quantization leads to wave functions given by the cohomology of $\IX_6$,
resulting in a net BPS multiplicity given by $\chi(\IX_6)$. 

In order to relate to type IIA, we compactify on an $\IS^1$.
Corrections to the $R^2$ term will arise from one-loop diagrams in which the above BPS particles run,
in the presence of the graviphoton field, which couples to their $\SU(2)_\L$ quantum numbers.
In type IIA language this corresponds to integrating out massive D0- and D2-brane states (and their bound states).
In the Schwinger proper time formalism we have
\beq
\begin{aligned}
\mathcal{F} &= \int_{\epsilon}^\infty \frac{\ud s}{s}
\tr_\hil \left[ (-1)^F e^{-s (\triangle + m^2 + J \cdot F_+) } \right]\\
&= \int_\epsilon^\infty \frac{\ud s}s \sum_{k\in \IZ}
\frac{1}{4 \sinh^2 (\frac {s\lambda}2)}
\tr_\hil \left[ (-1)^F e^{-2s \lambda J_3^L - sZ - 2\pi \iu s k} \right].
\end{aligned}
\label{gv-closed-n}
\eeq
Here the $\sinh^2$ factor arises from the 4d kinematics,
we have included a sum over KK momenta along the $\IS^1$,
the trace is over the Hilbert space $\hil$ of 5d one-particle BPS states,
with central charge $Z$,
and
$F=2J_3^L +2J_3^R$. 

The  Hilbert space $\hil$ of 5d one-particle BPS states from an M2-brane on a genus $g$ holomorphic curve $\Sigma_g$
(in general not the same as the genus of the worldsheet in the type IIA interpretation)
in the homology class $\beta$ is obtained by quantization of zero modes on its worldline.
Quantization of the universal Goldstinos contributes to the state transforming as a (half) hypermultiplet,
with $\SU(2)_\L$ representation $I_1=\im \frac12 \fm \oplus 2(0)$.
There are in general additional zero modes,
characterized in terms of the cohomology groups
\beq
 \hil = H^\bullet(\ms_{g,\beta}) \otimes H^\bullet(\IT^{2g}).
 \label{split}
\eeq
The first factor corresponds to zero modes from the deformation moduli space $\ms_{g,\beta}$ of $\Sigma_g$ in $\IX_6$,
whose quantization determines the $\SU(2)_\R$ representation.
The latter is decoupled from the self-dual graviphoton background,
so it only contributes as some extra overall multiplicity in the above trace.%
\footnote{This is no longer true if one considers refined topological strings as in \cite{Iqbal:2007ii,Gukov:2007tf,Aganagic:2011mi,Aganagic:2012au,Choi:2012jz},
corresponding to a non self-dual background field of the form
$F = \epsilon_1 \ud x^1 \wedge \ud x^2 - \epsilon_2 \ud x^3 \wedge \ud x^4$.}

The second factor corresponds to zero modes
arising from flat connections on the type IIA D2-brane worldvolume gauge field on $\Sigma_g$.
The $\IT^{2g}$ should be regarded as the Jacobian of $\Sigma_g$, $\jac \Sigma_g=\IT^{2g}$.
Quantization of these zero modes determines further contributions to the $\SU(2)_\L$ representation of the state
as dictated by the $\SU(2)$ Lefschetz decomposition of cohomology of $\IT^{2g}$,
i.e.\ with creation, annihilation and number operators
\beq 
J_+ = k \wedge, \quad J_-=k\lrcorner,\quad J_3=(\operatorname{deg}-n)/2.
\eeq
Here $k$ is the K\"ahler form of the torus,
$\lrcorner$ denotes contraction,
the bidegree $\operatorname{deg}$ is $p+q$ for a $(p,q)$-form
and $n$ is complex dimension.

The $\SU(2)$ representation is of the form $I_g = I_1^{\otimes g}$,
where $I_1=\im \frac12 \fm \oplus 2(0)$.
For instance, for $g=1$ we have a ground state 1 and operators $\ud z$ and $\ud \bz$,
so that the cohomology of $\IT^2$ splits as 
\beq
\im 1 \\ k \fm \quad \ud{z} \pm \ud \bz 
\label{lef}
\eeq
where $k \sim \ud z \wedge \ud \bz$.
These form the representation $I_1$.
The argument generalizes straightforwardly to higher genera.

The contribution from a state in the $\SU(2)_\L$ representation $I_g$ to the trace is given by
$(-4)^g \sinh^{2g} \frac {s\lambda}2$, so we get
\beq
\FF = - \int_\epsilon^\infty \frac{\ud s}s \sum_{g,\beta} \sum_{k\in \IZ}
\gv_{g,\beta} \left( 2\iu \sinh \frac {s\lambda}2 \right)^{2g-2} e^{-s \beta\cdot t} e^{-2\pi \iu s k},
\label{gv-interm}
\eeq
where we write $Z=\beta \cdot t$.
Also, $\gv_{g,\beta}$ are integers describing the multiplicity of BPS states
arising from M2-branes on a genus $g$ curve in the class $\beta\in H_2(\IX_6;\IZ)$, with the understanding that $\beta=0$ corresponds to 11d graviton states. This multiplicity includes that arising from the $\SU(2)_\R$ representations, in the following sense. Describing the set of BPS states in terms of their $\SU(2)_\L\times \SU(2)_\R$ representations
\beq \left[ \im \frac12 , 0 \fm \oplus 2(0,0) \right] \otimes \sum_{j_L,j_R} N^\beta_{j_L,j_R} (j_L,j_R)
\label{full-multiplet}
\eeq
we have
\beq \sum_{j_L,j_R} (-1)^{2j_R} (2j_R+1) N^\beta_{j_L,j_R} [j_L] = \sum_g \gv_{g,\beta} I_g.\eeq
Going back to \cref{gv-interm}, we use Poisson resummation
\beq \label{poisson}
\sum_{k\in \IZ} e^{-2\pi \iu sk} = \sum_{m\in \IZ} \delta(s-m)
\eeq
to get
\beq
\FF = - \sum_{g,\beta} \sum_{m=1}^\infty \gv_{g,\beta} \frac1m
\left( 2\iu \sinh \frac {m\lambda}2 \right)^{2g-2} e^{-m \beta\cdot t}.
\label{gv-closed-d}
\eeq
This is known as the GV or BPS expansion of the closed oriented topological string amplitude.
There are similar expansions for open oriented topological string amplitudes
(see \cref{sec:extra-pairs} for more details).
The situation for unoriented topological string amplitudes is the main topic of this paper.

\section{Walcher's real topological string}
\label{unoriented-gv}

A prominent example of unoriented topological string is Walcher's real topological string introduced in \cite{Walcher:2007qp}
(see also \cite{Krefl:2008sj,Krefl:2009md,Krefl:2009mw}).
It includes both open and closed unoriented topological strings,
subject to a mysterious `tadpole cancellation condition'
requiring the open string sector to be described by a single D-brane on top of the fixed locus of the orientifold action.
We now review its basic features, the proposed tadpole cancellation condition,
and the conjectured expansion in terms of integer multiplicities.
For other details, see \cref{sec:walcher}.
For simplicity, we take the case of a single K\"ahler modulus, although the generalization is straightforward.
Also, examples in the literature \cite{Walcher:2007qp,Krefl:2008sj,Krefl:2009md,Krefl:2009mw}
have considered cases with $H_1(L;\IZ) = \IZ_2$, for instance the quintic, or local $\CP^2$.
Since our description in M-theory is more general,
and this condition will only play a role in \cref{sec:tadpole},
we keep the description general here as well.

The A-model target space is a Calabi-Yau threefold $\IX_6$,
equipped with an antiholomorphic involution $\sigma$,
whose pointwise fixed set is a lagrangian 3-cycle denoted by $L$.
The model is defined by considering maps of (possibly non-orientable) surfaces (possibly with boundaries) into $\IX_6$,
with boundaries lying in $L$
(so that the maps are topologically classified
by the relative homology $d=f_*([\Sigma]) \in H_2(\IX_6,L;\IZ)$),
and with compatible orientifold actions on the target and worldsheet, as follows.
We construct the non-orientable surface $\Sigma$ as the quotient of the parent oriented Riemann surface $\hs$
by an antiholomorphic involution $\Omega$ of the worldsheet,%
\footnote{This is equivalent to considering a dianalytic structure on the surface,
which generalizes the notion of complex structure by allowing for antiholomorphic transition functions.}
and demand equivariance of the holomorphic map $f$ as in \cref{equivariance}.
\begin{figure}[ht]
\centering
\begin{tikzpicture}[node distance=2cm, auto]
  \node (P) at (-1,1) {$\hs$};
  \node (B) at (1,1) {$\hs$};
  \node (A) at (-1,-1) {$\IX_6$};
  \node (C) at (1,-1) {$\IX_6$}; 
  \node (D) at (0,0) {$\circlearrowleft$};
  \draw[->] (P) to node {$\Omega$} (B);
  \draw[->] (P) to node [swap] {$f$} (A);
  \draw[->] (A) to node [swap] {$\sigma$} (C);
  \draw[->] (B) to node {$f$} (C);
\end{tikzpicture}
\caption{Equivariance means the diagram is commutative
(we are improperly calling $f$ both the equivariant map and its lift $\Sigma \to X/\sigma$.)}
\label{equivariance}
\end{figure}

In the relation $\Sigma=\hs/\Omega$,
the particular case in which $\Sigma$ is itself closed oriented and $\hs$ has two connected components is not included.

The topological classification of possibly non-orientable surfaces $\Sigma$ with boundaries,
described as symmetric Riemann surfaces, written $(\hs,\Omega)$,
generalizes the closed oriented case, through the following classic result:
\begin{thm} 
Let $h=h(\hs,\Omega) \geq 0 $ be the number of components of the fixed point set $\hs_\Omega$ of $\Omega$ in $\hs$
\emph{(i.e.\ the number of boundaries of $\Sigma$)},
and introduce the index of orientability $k=k(\hs,\Omega)$,
given by $\left( 2 - \# \text{ components of }\hs \setminus \hs_\Omega \right)$. Then the topological invariants $h$ and $k$ together with the genus $\hg$ of $\hs$
determine the topological type of $(\hs,\Omega)$ uniquely.
For fixed genus $\hg$, these invariants satisfy
\begin{enumerate}
\item $k=0$ or $k=1$ (corresponding to oriented surfaces, or otherwise)
\item if $k=0$, then $0 < h \leq \hg+1$ and $h \equiv \hg+1 \mod{2}$
\item if $k=1$ then $h \leq \hg$.
\end{enumerate}
\end{thm}

Let us define the (negative of the) Euler characteristic of $\hs/\Omega$ by $\chi = \hg-1$.
It is useful to separate the worldsheets into three classes, corresponding to having 0, 1 or 2 crosscaps
(recall that two crosscaps are equivalent to a Klein handle,
namely two holes glued together with an orientation reversal,
which in the presence of a third crosscap can be turned into an ordinary handle).
This leads to a split of the topological amplitudes into classes, namely:
closed oriented surfaces
(with amplitude denoted by $\mathcal F^{(g_\chi)}$, with $g_\chi=\frac12 \chi + 1$ the number of handles),
oriented surfaces with $h$ boundaries
(with amplitude $\mathcal F^{(g,h)}$),
non-orientable surfaces with an odd number of crosscaps
(with amplitude $\mathcal R^{(g,h)}$)
and non-orientable surfaces with an even number of crosscaps
($\mathcal K^{(g,h)}$).
The Euler characteristic is given by $\chi = 2g -2 +h +c$, with $c=0,1$.%
\footnote{Note that we define $g$ such that the negative Euler characteristic is
$2g+h-2$, $2g+h-1$, and $2g+h-2$ in the $\mathcal{F}$, $\mathcal{R}$ and $\mathcal{K}$ cases respectively,
i.e.\ it also accounts for Klein handles.}

\bigskip

The basic tool used to compute these amplitudes is equivariant localization on the moduli space $\ms$ of stable maps,
following ideas going back to \cite{Kontsevich:1994na}
(see also \cite{Diaconescu:2003dq}, and \cite{mathematicians,categories} for more recent developments on the formal side).
Localization is with respect to a torus action which is compatible with the involution,
and leads to a formulation in terms of the diagram techniques of \cite{Kontsevich:1994na}.

Ref.~\cite{Walcher:2007qp} finds that, in the example of the quintic or local $\CP^2$,
in order to apply this machinery to unoriented and/or open worldsheets,
some constraints, dubbed \emph{tadpole cancellation} conditions, have to be imposed:
as we discuss in \cref{sec:walcher},
this is a cancellation between contributions from worldsheets with an unpaired crosscap and worldsheets with boundaries,
with one boundary ending on $L$ `with even degree'
(specifically, wrapping the generator of $H_1(L,\IZ)=\IZ_2$ an even number of times, hence begin topologically trivial).
This results in a condition
\beq
\label{tadpole-canc} 
d \equiv h \equiv \chi \mod{2},
\eeq
where $d \in H_2(\IX,L;\IZ)=\IZ$ is the relevant homology class.
It implies that $\mathcal{R}$-type amplitudes do not contribute, $\mathcal R^{(g,h)}\equiv 0$.

Mathematically, this condition applies to real codimension one boundary strata in moduli space,
in which a given worldsheet piece near $L$ develops a node which can be smoothed to yield
either a disk or a crosscap.
The combined count of these homologically trivial disks and crosscaps
leads to cancellation of potentially ill-defined pieces,
and produces an invariant count.

\bigskip

Strong evidence for this consistency condition comes from the fact that the invariant numbers thus computed turn out to be all integers.
This motivated the proposal of an ansatz reminiscent of a BPS expansion,
as a sum over holomorphic \emph{embeddings} (rather than maps) equivariant with respect to worldsheet parity $\Omega:\hs \to \hs$.

If we write the total topological amplitude as
\beq
\mathcal{G}^{(\chi)} =
\frac12 \left[ \mathcal{F}^{(g_\chi)} + \sum \mathcal{F}^{(g,h)} + \sum \mathcal{K}^{(g,h)} \right],
\label{total-g}
\eeq
the conjecture is
\beq
\label{walcher-conj} 
\sum_{\chi} \iu^\chi \lambda^\chi \left( \mathcal{G}^{(\chi)} - \frac12 \mathcal{F}^{(g_\chi)} \right) =
\sum_{\substack{\chi \equiv d \bmod{2} \\ k \text{ odd}}}
\tgv_{\hat g,d} \frac1k \left( 2 \sinh \frac{\lambda k}2 \right)^\chi q^{kd/2}.
\eeq
In more physical terms, the tadpole cancellation condition means that the background contains a single D-brane wrapped on $L$,
as counted in the covering space.
The interpretation in terms of a physical type IIA construction and its lift to M-theory will be discussed in the next section.

\section{M-theory lift and BPS expansion}
\label{sec:mth}

\subsection{Tadpole cancellation, the O4/D4 system and M-theory}
\label{sec:lift}

It is natural to look for a physical realization of the real topological string in terms of type IIA on the threefold $\IX_6$,
quotiented by worldsheet parity times an involution acting antiholomorphically on $\IX_6$.
In general, we consider involutions with a fixed point set along the lagrangian 3-cycle $L$, which therefore supports an orientifold plane.
The total dimension of the orientifold plane depends on the orientifold action in the 4d spacetime,
and can correspond to an O6-plane or an O4-plane.
The choice of 4d action is not specified in the topological string, but can be guessed as follows.

We expect that the topological tadpole cancellation condition has some translation in the physical theory,
as a special property occurring when precisely one D-brane (as counted in the covering space) is placed on top of the orientifold plane.
Since the charge of a negatively charged O4-plane is $-1$ (in units of D4-brane charge in the covering),
the configuration with a single D4-brane stuck on top of it is special,
because it cancels the RR charge locally (on the other hand, the charge of an O6-plane is $-4$, and no similarly special property occurs for a single stuck D6-brane).

The presence of a single D4-brane stuck on the O4-plane is not
a consistency requirement of the type IIA theory configuration,%
\footnote{Notice that any configuration with additional pairs of D4-branes is continuously connected to it.
See \cref{sec:extra-pairs} for further discussion.
Also, topological A-models related to systems of O4-planes with no stuck D4-brane, and their M-theory interpretation,
have appeared in \cite{Aganagic:2012au},
see \cref{sec:others} for further discussion.}
but rather a condition that we will show leads to a particularly simple M-theory lift,
and a simple extension of the Gopakumar-Vafa BPS expansion of topological amplitudes.
This nicely dovetails the role played by tadpole cancellation
in the real topological string to achieve the appearance of integer invariants.

The M-theory lift of O4-planes with and without D4-branes has been discussed in \cite{Hori:1998iv,Gimon:1998be}.
In particular, a negatively charged O4-plane with no stuck D4-brane, spanning the directions 01234 in 10d Minkowski space $M_{10}$,
lifts to M-theory on a $\IZ_2$ orbifold $M_5\times (\IR^5/\IZ_2) \times \IS^1$ with generator $(x^5,\ldots,x^{9})\to (-x^5,\ldots,-x^9)$,
and which also flips the M-theory 3-form, $C_3\to -C_3$. The latter action is required to be a symmetry of the M-theory Chern-Simons term,
and matches the effect of the type IIA orientifold action on the NSNS 2-form $B_2$.
Hence, we will classify the M-theory action as `orientifold' as well.

The construction generalizes to compactification on $\IX_6$,
with the orientifold acting holomorphically on $\IX_ 6$.
It produces M-theory on the quotient $(M_4 \times {\IX_6})/\IZ_2 \times \IS^1$,
with the $\IZ_2$ acting as the antiholomorphic involution $\sigma$ on $\IX_6$ and as $x^{2},x^3\to -x^2,-x^3$ on 4d Minkowski space.
This system, and its generalization with additional D4-brane pairs (M5-branes in M-theory), is discussed in \cref{sec:others}.
Here we simply note that the explicit breaking of the $\SU(2)_\L$ symmetry already in the 5d theory makes necessary
to make certain assumptions on the structure of BPS multiplets in the theory,
obscuring the derivation of the BPS expansion of the amplitude.

The M-theory lift of a negatively charged O4-plane with a stuck D4-brane is however much simpler,
and in particular does not suffer from these difficulties.
Because of the already mentioned local cancellation of the RR charge,
the M-theory lift is a completely smooth space described by a freely acting quotient
$M_5\times (\IR^5\times \IS^1)/\IZ_2$,
with generator acting as $(x^5,\ldots,x^{9})\to (-x^5,\ldots,-x^9)$,
as a half-period shift on $\IS^1$ ($y\to y+\pi$ for periodicity $y\simeq y+2\pi$),
and flipping the 3-form $C_3$ (hence defining an M-theory orientifold).

Concerning the latter,
it is important to point out that the negative charge of the O4-plane implies that
there is a half-unit NSNS $B_2$ background on an $\RP^2$ surrounding the O4-plane;
consequently, there is a non-trivial half-unit of 3-form background on the corresponding M-theory lift $(\CP^1\times\IS^1)/\IZ_2$.
This will play an important role in the M-theory interpretation of the disk/crosscap tadpole cancellation,
see \cref{sec:tadpole}.

The construction generalizes to compactification on $\IX_6$,
with the orientifold acting holomorphically on $\IX_ 6$.
It produces M-theory on the quotient $(M_4 \times {\IX_6} \times \IS^1)/\IZ_2$,
with the $\IZ_2$ acting as
\beq \begin{aligned}
\IX_6: \quad x \mapsto \sigma(x), \qquad \IS^1: \quad  y \mapsto y + \pi,\\
\text{Minkowski}: \quad x^0,x^1 \mapsto x^0,x^1, \qquad x^2,x^3 \mapsto - x^2,-x^3.\end{aligned}
\eeq
The geometry is a (M\"obius) fiber bundle
with base $\IS^1$, fiber $M_4 \times \IX_6$, and structure group $\IZ_2$.

As before, it is straightforward to add extra D4-brane pairs away from (or on top of) the O4-plane,
since they lift to extra M5-brane pairs in M-theory,
see \cref{sec:extra-pairs}.

\subsection{M-theory BPS expansion of the real topological string}
\label{sec:bps}

The M-theory configuration allows for a simple Gopakumar-Vafa picture of amplitudes,
which should reproduce the real topological string amplitudes.
Since the quotient is acting on the M-theory $\IS^1$ as a half-shift, its effect is not visible locally on the $\IS^1$.
This means that the relevant 5d picture is exactly the same as for the closed oriented setup, c.f.\ \cref{sec:gv}, so
the relevant BPS states are counted by the standard Gopakumar-Vafa invariants.
When compactifying on $\IS^1$ and quotienting by $\IZ_2$, these states run in the loop as usual,
with the only (but crucial) difference that they split according to their parity under the M-theory orientifold action.
In the M\"obius bundle picture, even components of the original $\NN=2$ multiplets will run on $\IS^1$ with integer KK momentum,
whereas odd components run with half-integer KK momentum. 

The split is also in agreement with the reduction of supersymmetry by the orientifold, which only preserves 4 supercharges.
Note also that the orientifold is not 4d Poincar\'e invariant, as Lorentz group is broken as
\beq 
\SO(4)=\SU(2)_\L \times \SU(2)_\R \to \U(1)_\L \times \U(1)_\R.
\eeq
The preserved supersymmetry is not 4d $\NN=1$ SUSY, and in particular it admits BPS particles. 

\subsubsection{General structure}

The states in the Hilbert space $\hil$ are groundstates in the SUSY quantum mechanics on the moduli space of wrapped M2-branes.
In the orientifold model, these BPS states of the 5d theory fall into two broad classes.

\subsubsection*{Non-invariant states and the closed oriented contribution}

Consider a BPS state $|A\rangle$ associated to an M2 on a curve $\Sigma_g$ not mapped to itself under the involution $\sigma$;
there is an image multiplet $|A'\rangle$ associated to the image curve%
\footnote{Actually, the BPS state wavefunctions may be partially supported in the locus in moduli space where the curve and its image combine into an irreducible equivariant curve.
Hence, part of the contribution of non-invariant states in this section may spill off to contributions of invariant states discussed later.\label{spill}}
$\Sigma'_g$.
We can now form orientifold even and odd combinations $|A\rangle \pm |A'\rangle$,
which run on the $\IS^1$ with integer or half-integer KK momentum, respectively.
Since each such pair has $Z_A=Z_{A'}$ and identical multiplet $\SU(2)_\L \times \SU(2)_\R$ structure and multiplicities
(inherited from the parent theory),
we get the following structure:
\begin{align}
\begin{split}
 & \int_\epsilon^\infty \frac{\ud s}s \sum_{k\in \IZ} \frac{1}{4\sinh^2 (\frac {s\lambda}2)} e^{-sZ} \tr
 \Big\{ 
   \!\begin{aligned}[t] & \left[ (-1)^{F} e^{-2s\lambda J_3^L-2\pi \iu sk} \right] \\
                    & + \left[ (-1)^{F} e^{-2s\lambda J_3^L - 2\pi \iu s (k+\frac 12)} \right] \Big\}
      \end{aligned}
     \\
&=\int_\epsilon^\infty \frac{\ud s}s \frac{1}{4\sinh^2 (\frac {s\lambda}2)}
e^{-sZ} \tr
\left[ (-1)^{2J_3^L+2J_3^R} e^{-2s\lambda J_3^L} \right]
\sum_{k \in \IZ} \left( e^{-2\pi \iu s k} + e^{-2\pi \iu s(k+\frac 12)} \right) \\
& = -2\int_\epsilon^\infty \frac{\ud s}s n_{\Sigma_g}
\left( 2 \iu\sinh \frac {s\lambda}2 \right)^{2g-2} e^{-sZ} \sum_m \delta(s-2m) \\
& = -2\sum_{\text{even } p>0} n_{\Sigma_g} \frac{1}{p}
\left( 2\iu \sinh \frac{p\lambda}2 \right)^{2g-2} e^{-p Z}.
\end{split}
\end{align}
The $\sinh^{-2}$ factor corresponds to 4d kinematics, since the orientifold imposes no restriction on momentum in the directions transverse to the fixed locus.
We have also denoted $n_{\Sigma_g}$ the possible multiplicity arising from $\SU(2)_\R$ quantum numbers.
Clearly, because the states are precisely those in the parent $\NN=2$ theory,
summing over multiplets reproduces the Gopakumar-Vafa expansion \cref{gv-closed-d} of the closed oriented contribution
to the topological string partition function:
this is because even wrappings $p$ on the orientifold $\IS^1$
correspond to both even and odd wrappings on the closed-oriented $\IS^1$, due to the reduction of the $\IS^1$ by half.
This closed oriented contribution must be duly subtracted from the total amplitude, in order to extract the genuine contribution associated to equivariant curves, reproducing the open and unoriented piece of the topological amplitude;
this nicely reproduces the subtraction of the closed oriented contribution in the left hand side of \cref{walcher-conj}.

The conclusion is that contributions with even wrapping belong to the sector of non-invariant states, which heuristically describe disconnected curves in the cover and reproduce the closed oriented topological string.

\subsubsection*{Invariant states and the open and unoriented contributions}

The second kind of BPS states correspond to M2-branes wrapped on curves $\hs_{\hg}$ in the cover,
mapped to themselves under $\sigma$.
The overall parity of one such state
is determined by the parities of the states in the corresponding $\SU(2)_\L$ and $\SU(2)_\R$ representations.
We introduce the spaces $\hil^\hg_\pm$ describing the even/odd pieces of the $\SU(2)_\L$ representation $I_\hg$ for fixed $\hg$. In what follows, we drop the $\hg$ label to avoid cluttering notation. We similarly split the equivariant BPS invariant 
$\gv'_{\hg,\beta}$
(i.e.\ after removing the pairs of states considered in the previous discussion)
into even/odd contributions as
\beq
\gv'_{\hg,\beta} = \gvp + \gvm.
\label{gv-parity}
\eeq
Recalling that states with even/odd overall parity have integer/half-integer KK momenta, we have a structure
\begin{align}
\begin{split}
& \sum_{\beta,\hg} e^{-sZ}
\Big[
\!\begin{aligned}[t]
& \left( \gvp \tr_{\hil_+} \op + \gvm \tr_{\hil_-} \op \right)
\sum_{k\in \IZ} e^{-2\pi \iu s k}\\
& + \left( \gvp \tr_{\hil_-} \op + \gvm \tr_{\hil_+} \op \right)
\sum_{k\in \IZ} e^{-2\pi \iu s (k+\frac 12)} \Big]\end{aligned}\\
& = \sum_{\beta,\hg} e^{-sZ} \sum_{m\in \IZ} \delta(s-m)
\Big\{
\!\begin{aligned}[t]
& \left[ \gvp \tr_{\hil_+} \op + \gvm \tr_{\hil_-} \op \right]\\
& +(-1)^m \left[\gvp \tr_{\hil_-} \op
+ \gvm \tr_{\hil_+} \op \right] \Big\} \end{aligned}\\
& = \sum_{\beta,\hg} e^{-sZ} \sum_{m\in 2\IZ} \delta(s-m)\,
\left( \gvp+\gvm\right)\,
\left\{ \tr_{\hil_+} \op + \tr_{\hil_-} \op \right\} \\
& \quad + \sum_{\beta,\hg} e^{-sZ} \sum_{m\in 2\IZ+1} \delta(s-m)\,
\left( \gvp-\gvm \right)\,
\left\{ \tr_{\hil_+} \op - \tr_{\hil_-} \op \right\}.
\label{mess}
\end{split}
\end{align}
In the next to last line,
the traces clearly add up to the total trace over the parent $\NN=2$ multiplet
and the $\gv'^\pm$ add up to the parent BPS invariants, c.f.\ \cref{gv-parity}.
Noticing also that it corresponds to even wrapping contributions $m\in 2\IZ$,
we realize that this corresponds to a contribution to the closed oriented topological string partition function,
c.f.\ \cref{spill}.
As discussed, it should not be included in the computation leading to the open and unoriented contributions.

The complete expression for the latter is
\beq
\begin{aligned}
& \sum_{\mathclap{\substack{\beta,\hg\\ m \in 2\IZ+1} }}\;
\hgv
\int_\epsilon^\infty \frac{\ud s}s \frac{ \delta(s-m)}{2\sinh (\frac {s\lambda}2)}
e^{-sZ} \left\{ \tr_{\hil_+} \left[ (-1)^{2J_3^L} e^{-s\lambda J_3^L} \right] - \tr_{\hil_-} \left[ (-1)^{2J_3^L} e^{-s\lambda J_3^L} \right] \right\} \\
&=
\sum_{\mathclap{\substack {\beta,\hg \\ \text{odd } m \geq 1} }}\;
\hgv \frac{1}{m} \frac{1}{2\sinh (\frac {m\lambda}2)}
e^{-mZ} \left\{ \tr_{\hil_+} \left[ (-1)^{2J_3^L} e^{-m\lambda J_3^L} \right] - \tr_{\hil_-} \left[ (-1)^{2J_3^L} e^{-m\lambda J_3^L} \right] \right\},
\end{aligned}
\label{semi-final-trace}
\eeq
where we have introduced the integers, which we call real BPS invariants,
\beq
\hgv := \gvp - \gvm.
\label{real-gv}
\eeq
These integer numbers $\hgvs$'s are those playing the role $\tgv$'s in \cref{walcher-conj}.
Note however that their correct physical interpretation differs from that in \cite{Walcher:2007qp},
where they were rather identified as our $\gv'_{\hg,\beta}$.
Note also that the correct invariants \cref{real-gv} are equal mod 2 to the parent $\gv_{\hg,\beta}$,
proposed in \cite{Walcher:2007qp}, just like the $\gv'_{\hg,\beta}$.

In \cref{semi-final-trace} we have taken into account that these states,
being invariant under the orientifold, propagate only in the 2d fixed subspace of the 4d spacetime,
resulting in a single power of $(2\sinh)$ in the denominator.
This also explains the factor of 2 in the graviphoton coupling relative to \cref{gv-closed-n}.
In the next section we fill the gap of showing the promised equality of the even and odd multiplicities,
and compute the trace difference in the last expression.

\subsubsection{Jacobian and computation of \texorpdfstring{$\SU(2)_\L$}{SU(2)L} traces}

We must now evaluate the trace over the even/odd components of the Hilbert space of a parent $\NN=2$ BPS multiplet.
This is determined by the parity of the corresponding zero modes on the particle worldline.
As reviewed in \cref{sec:gv},
the traces are non-trivial only over the cohomology of the Jacobian of $\hs_\hg$ which determines the $\SU(2)_\L$ representation. We now focus on its parity under the orientifold. 

Consider for example the case of $I_1$, c.f.\ \cref{lef}.
We introduce the formal split of the trace into traces over $\hil_\pm$
\beq
t_1 = t_1^+ \ominus t_1^- 
\label{split-tracce1},
\eeq
where $\pm$ denotes orientifold behavior and $\ominus$ denotes a formal combination operation,
which satisfies $\ominus^2=1$
(it corresponds to the $(-1)^m$ factor once the wrapping number $m$ has been introduced, c.f.\ \cref{mess}).
Since the orientifold action is an antiholomorphic involution on the worldsheet,
it acts as $\ud z \leftrightarrow \ud\bz$, so \cref{lef} splits as
\beq
I_1 = \im \frac12 \fm \oplus 2 \im 0 \fm = \im + \\ - \fm \oplus \im + \fm \oplus \im - \fm,
\eeq
which gives a trace
\beq 
t_1 = \underbrace{(1-e^{s/2})}_{t_1^+} \ominus \underbrace{(1-e^{-s/2})}_{t^-_1},
\label{lef-uno}
\eeq
where, to avoid notational clutter, we have reabsorbed $\lambda$ into $s$.

Since the creation and annihilation operators associated to different 1-forms commute,
the argument generalizes easily to higher genus, and the trace over a representation $I_\hg$ has the structure
\beq 
t_\hg = (t_1^+ \ominus t_1^-)^\hg = t_\hg^+ \ominus t_\hg^-, 
\label{split-tracce}
\eeq
where $t_\hg^+$ and $t_\hg^-$ contain even and odd powers of $t_1^-$, respectively.
For instance, for $I_2$ we have to trace over
\beq \im + \\ - \\ + \fm \oplus 2 \im + \\ - \fm \oplus 2 \im - \\ + \fm \oplus 2 \im + \fm \oplus 3 \im - \fm\eeq
and obtain
\beq
t_2 =
\underbrace{(t^+_1)^2 + (t_1^-)^2}_{2 + e^{-s} + e^{s} -2e^{-s/2} -2e^{s/2}} \ominus \underbrace{2 t_1^+ t_1^-}_{4-2e^{-s/2} -2e^{s/2}}
=(t_1^+ \ominus t_1^-)^2 = t_2^+ \ominus t_2^-.
\label{lef-uno2}
\eeq
We are now ready to compute the final expression for the BPS expansion

\subsubsection{The BPS expansion}

Recalling \cref{walcher-conj},
the genuine open and unoriented contribution reduces to the odd wrapping number case \cref{semi-final-trace}.
Interestingly, the trace difference can be written (restoring the $\lambda$)
\beq
t_\hg^+ - t_\hg^- = \left( -2\sinh \frac {s\lambda}2 \right)^\hg.
\eeq
This is clear from \cref{lef-uno,lef-uno2} for $I_1$, $I_2$ respectively, and holds in general.

The final result for the BPS amplitude, which corresponds to the BPS expansion of the open and unoriented partition function, is
\beq
\sum_{\substack {\beta,\hg \\ \text{odd } m \geq 1} } \hgv \, \frac1m
\left[ \, 2\sinh \left( \frac {m\lambda}2 \right) \,\right] ^{\hg-1}
e^{-mZ}.
\label{final-trace}
\eeq
This has the precise structure to reproduce the conjecture in \cite{Walcher:2007qp} as in \cref{walcher-conj},
with the invariants defined by \cref{real-gv}.
In particular we emphasize the nice matching of exponents of the $\sinh$ factors (achieved since for the covering $\hg-1 = \chi$)
and of the exponential $e^{-Z}=q^{d/2}$ for a one-modulus $\IX_6$
(the factor of $1/2$ coming from the volume reduction due to the $\IZ_2$ quotient.)

The only additional ingredient present in \cref{walcher-conj} is the restriction on the degree,
which is related to the conjectured tadpole cancellation condition,
and which also admits a natural interpretation from the M-theory picture, as we show in the next section.
We simply advance that this restriction applies to examples with $H_1(L;\IZ)=\IZ_2$.
Our formula above is the general BPS expansion of the real topological string on a general CY threefold.

\bigskip

We anticipate that, once the tadpole cancellation discussed below is enforced,
our derivation of \cref{final-trace} provides the M-theory interpretation
for the integer quantities $\tgv$ appearing in \cref{walcher-conj}
as conjectured in \cite{Walcher:2007qp}.
Therefore the real topological string is computing
(weighted) BPS multiplicities of equivariant M2-brane states in M-theory,
with the weight given by an orientifold parity sign, c.f.\ \cref{real-gv}.
It would be interesting to perform a computation
of the numbers appearing in \cref{final-trace}
along the lines of \cite{Huang:2006hq,Katz:1999xq}.

\subsection{Tadpole cancellation}
\label{sec:tadpole}

In this section we discuss the M-theory description of the tadpole cancellation condition,
in examples of the kind considered in the literature,
i.e.\ with $H_1(L;\IZ)=\IZ_2$ and $H_2(\IX_6,L;\IZ)=\IZ$ (like the quintic or local $\CP^2$), for which one trades the class $\beta$ for the degree $d\in\IZ$.
The argument involves several steps.

\subsubsection*{First step: Restriction to even degree}

Consider the relative homology exact sequence
\beq
\equalto{H_2(\IX_6;\IZ)}{\IZ} \stackrel{2\cdot}{\to} \equalto{H_2(\IX_6,L;\IZ)}{\IZ}
\to \equalto{H_1(L;\IZ)}{\IZ_2}.
\eeq
Since (the embedded image of) a crosscap doesn't intersect the lagrangian $L$,
its class must be in the kernel of the second map, i.e.\ the image of the first.
Thus, every crosscap contributes an even factor to the degree $d$.
For boundaries,
the same argument implies that boundaries wrapped on an odd multiple of the non-trivial generator of $H_1(L;\IZ)=\IZ_2$
contribute to odd degree,
while those wrapped on an even multiple of the $\IZ_2$ 1-cycle contribute to even degree.
This restricts the possible cancellations of crosscaps to even degree boundaries.

\subsubsection*{Second step: Relative signs from background form fields} 

We now show that there is a relative minus sign between crosscaps and disks associated to the same (necessarily even degree) homology class. As mentioned in \cref{sec:lift},
the M-theory lift contains a background 3-form $C_3$ along the 3-cycles $(\CP^1 \times \IS^1)/\IZ_2$,
with the $\IZ_2$ acting antiholomorphically over $\CP^1$;
this corresponds to a half-unit of NSNS 2-form flux on any crosscap $\RP^2$ surrounding the O4-plane in the type IIA picture.
In M-theory, the reduction of $C_3$ along the $\CP^1$ produces a 5d gauge boson,
under which any M2-brane is charged with charge $c$,
where $c$ is the number of crosscaps in the embedded curve in $\IX_6/\sigma$.
The 3-form background corresponds to a non-trivial $\IZ_2$ Wilson line turned on along the M-theory $\IS^1$, and produces an additional contribution to the central charge term $Z$, which (besides the KK term) reads  
\beq 
Z = dt + \frac{\iu}2 c.
\eeq
Once we exponentiate, and perform Poisson resummation \cref{poisson},
this gives a contribution $(-1)^{m \times c}$, with $m$ the wrapping number,
which is odd for the genuine equivariant contributions.
This extra sign does not change the contributions of curves with even number of crosscaps.
Note that the above also agrees with the fact that the (positive) number of crosscaps is only defined mod 2.

In contrast,  boundaries do not receive such contribution,%
\footnote{Note that odd degree boundaries can receive an extra sign due to a possible $\IZ_2$ Wilson line on the D4-brane.
This however does not affect the even degree boundaries, which are those canceling against crosscaps.
Hence it does not have any effect in the explanation of the tadpole cancellation condition.}
and therefore there is a relative sign between contributions from curves which fall in the same homology class,
but differ in trading a crosscap for a boundary. 

\subsubsection*{Third step: Bijection between crosscaps and boundaries} 

To complete the argument for the tadpole cancellation condition, one needs to show that there is a one-to-one correspondence between curves which agree except for a replacement of one crosscap by one boundary. The replacement can be regarded as a local operation on the curve, so the correspondence is a bijection between disk and crosscap contributions. 

More precisely we want to show that for every homologically trivial disk
which develops a node on top of $L$ we can find a crosscap, and viceversa.
This is mathematically a nontrivial statement,
for which we weren't able to find an explicit construction going beyond the local model of \cref{veronese}.
Moreover, in higher genera this problem has not been tackled by mathematicians yet.
Nonetheless, following \cite{tehrani}, we propose a model for gluing boundaries on the moduli space
in the genus zero case,
which points towards the desired bijection. The original argument applies to holomorphic maps, relevant to Gromov-Witten invariants; we expect similar results for holomorphic embeddings, relevant for Gopakumar-Vafa invariants.

The main point is that integrals over moduli spaces of Riemann surfaces make sense
and are independent of the choice of complex structure if the moduli space has a virtually orientable fundamental cycle
without real codimension one boundaries (RCOB).
If $L \neq \emptyset$, in order to achieve this,
one has to consider together
contributions coming roughly speaking from open and unoriented worldsheets,
as proposed by \cite{Walcher:2007qp}.

We are interested in elements of RCOB in which a piece of the curve degenerates as two spheres touching at a point $q$:
\beq 
(f,\Sigma = \Sigma_1 \cup_q \Sigma_2), 
\eeq
where $f$ is the holomorphic map, $\Sigma_i = \CP^1$, the involution exchanges the $\Sigma_i$ and $f(q)\in L$.
For real $\epsilon \neq 0$ one can glue $\Sigma$ into a family of smooth curves, described locally as%
\footnote{In this equivariant covering picture,
we only consider singularities of type (1) as in Definition 3.4 of \cite{liu},
since the covering does not have boundaries.}
\beq 
\Sigma_\epsilon = \left\{ (z,w) \in \IC^2 : \quad zw = \epsilon \right\}.
\eeq
For $\epsilon \in \IR$, $\Sigma_\epsilon$ inherits complex conjugation from \cref{involutions},
and the fixed point set is $\IS^1$ if $\epsilon > 0$, empty if $\epsilon < 0$.
They correspond to an equivariant curve with two different involutions,
which in terms of homogeneous coordinates on $\CP^1$, can be described as
\beq (u:v) \mapsto (\ov{u}: \pm \ov{v}),
\label{involutions}
\eeq
leading to either a boundary or a crosscap.
The RCOB corresponding to sphere bubbling in the `$+$' case
is the same as the RCOB for the `$-$' case.
By attaching them along their common boundary,
we obtain a moduli space whose only RCOB
corresponds to disk bubbling. The resulting combined moduli space admits a Kuranishi structure and produces well-defined integrals.

\subsubsection*{Final step}

Using the above arguments, we can now derive \cref{tadpole-canc}, as follows.
First, the tadpole cancellation removes contributions where the number of crosscaps $c$ is odd,
so taking $\chi = 2g -2 +h +c$ we have $\chi \equiv h \pmod{2}$,
where $h$ denotes the number of boundaries.
Second, the value of $d=\sum d_i \pmod {2}$
can only get contributions from boundaries and crosscaps
(since contributions of pieces of the curve away from $L$ cancel mod 2 from the doubling due to the orientifold image);
moreover contributions from crosscaps and even degree boundaries cancel.
Hence, the only contributions arise from boundaries with odd terms $d_i$,
so clearly $d \equiv h \pmod{2}$.
We hence recover \cref{tadpole-canc}.

\section{Extensions and relations to other approaches}
\label{sec:extensions}

\subsection{Adding extra D4-brane pairs}
\label{sec:extra-pairs}

The discussion in the previous sections admits simple generalizations,
for instance the addition of $N$ extra D4-brane pairs in the type IIA picture.%
\footnote
{The addition of one extra unpaired D4-brane would modify drastically the M-theory lift of the configuration,
as discussed in the next section.}
The two branes in each pair are related by the orientifold projection,
but can otherwise be placed at any location, and wrapping general lagrangian 3-cycles in $\IX_6$.
For simplicity, we consider them to wrap the O4-plane lagrangian 3-cycle $L$,
and locate them on top of the O4-plane in the spacetime dimensions as well.

In the M-theory lift, we have the same quotient acting as a half shift on the M-theory $\IS^1$
(times the antiholomorphic involution of $\IX_6$ and the spacetime action $x^2,x^3\to -x^2,-x^3$),
now with the extra D4-brane pairs corresponding to extra M5-brane pairs,
related by the M-theory orientifold symmetry \cite{Hori:1998iv}.
Notice that since the orientifold generator is freely acting,
there is no singularity, and therefore no problem in understanding the physics associated to these M5-branes.
In the 5d theory, there is no orientifold,
and the introduction of the M5-branes simply introduces sectors of BPS states
corresponding to open M2-branes wrapped on holomorphic 2-chains with boundary on the M5-brane lagrangians. Their multiplicity is precisely given by the
open oriented Gopakumar-Vafa invariants \cite{Ooguri:1999bv}. These particles run in the M-theory $\IS^1$ and  must be split according to their parity under the orientifold action,
which determine the appropriate KK momentum quantization.
By the same arguments as in \cref{sec:bps}, the contributions which have even wrapping upon Poisson resummation actually belong to the open oriented topological amplitude, and should be discarded. To extract the genuine open unoriented amplitude, we must focus on surfaces mapped to themselves under $\sigma$, and restrict to odd wrapping number.
In analogy with \cref{sec:bps} and \cite{Ooguri:1999bv}, the amplitude can be written
\beq
\sum_{\text{odd } m \geq 1} \sum_{\beta,r,\cal {R}}
\frac{N^+_{\beta,r, {\cal{R}}} - N^-_{\beta,r,{\cal{R}}}} {2m\sinh(\frac{m\lambda}2)}
e^{-m\beta\cdot t - mr\lambda} \tr_{\cal{R}} \prod_{i=1}^{b_1(L)} V_i^m.
\label{final-trace-open}
\eeq
The sum in $m$ runs only over positive odd integers.
The $N^{\pm}_{\beta,{\cal{R}},r}$ are the multiplicities of (even or odd) states from M2-branes on surfaces in the class $\beta$,
with spin $r$ under the rotational $\U(1)$ in the 01 dimensions,%
\footnote{Actually, it is the charge under the $\U(1)_\L\subset \SU(2)_\L$, which describes the coupling to the self-dual graviphoton background.}
and in the representation ${\cal{R}}$ of the background brane $\SO(2N)$ symmetry. The $V_i$ denote the lagrangian moduli describing the $\SO(2N)$ Wilson lines (complexified with deformation moduli), possibly turned on along the non-trivial 1-cycles of $L$. Since the presence of the M5-branes breaks the $\SU(2)_\L$ structure, it is not possible to perform a partial sum over such multiplets explicitly.

A clear expectation from the type IIA perspective is that the total amplitude (namely adding the original contribution in the absence of D4-brane pairs) could be rewritten to display an $\SO(2N+1)$ symmetry, combining the stuck and paired D4-branes.  This is possible thanks to the close analogy of the above expression with \cref{final-trace},
once we expand the contribution $\sinh^{\hg}$ to break down the $\SU(2)_\L$ multiplet structure.
By suitably subtracting contributions $N^+-N^-$ to the multiplicities $\hat \gv$,
one can expect to isolate the $\SO(2N+1)$ symmetric contribution.

For instance, take the case of only one matrix $V$ (e.g.\ $H_1(L;\IZ)=\IZ$ or $\IZ_2$). Combining \cref{final-trace-open} with \cref{final-trace}, we have
\beq
\sum_{\beta,r,m} \frac{e^{-mZ}}{2m\sinh(\frac{m\lambda}{2})} e^{-mr\lambda}\,
\left[ \hgvs_{r,\beta}  + \left( N^+_{\beta,r,{\cal{R}}} - N^-_{\beta,r,{\cal{R}}} \right)
\tr_{\cal {R}} V^m \right],
\eeq
where we have introduced $\hat \gv_{r,\beta}$ as the combination of real BPS invariants $\hgv$
describing the multiplicity of M2-brane states with 2d $\U(1)$ spin $r$.
For a given representation ${\cal R}$, the requirement that the expression in square brackets combines into traces of $\SO(2N+1)$ seems to imply non-trivial relations between the open and real BPS invariants for different representations of $\SO(2N)$. It would be interesting to study these relations further.

\subsection{Relation to other approaches}
\label{sec:others}

In this section we describe the relation of our system with other unoriented A-model topological strings
and their physical realization in M-theory.

\subsubsection{M-theory lift of the four O4-planes}

As discussed in \cite{Hori:1998iv,Gimon:1998be} there are four kinds of O4-planes in type IIA string theory, 
with different lifts to M-theory. We describe them in orientifolds of type IIA on $\IX_6\times M_4$,
with the geometric part of the orientifold acting as an antiholomorphic involution on $\IX_6$ and $x^2,x^3\to -x^2,-x^3$ on the 4d spacetime.%
\begin{itemize}
\item An O4$^-$-plane
(carrying $-1$ units of D4-brane charge, as counted in the covering space)
with no D4-branes on top.
Its lift to M-theory is a geometric orientifold $M_2\times (\IR^2\times \IX_6)/\IZ_2\times \IS^1$.
Inclusion of additional D4-brane pairs (in the covering space) corresponds to including additional M5-brane pairs in the M-theory lift.

\item An O4$^0$-plane, which can be regarded as an O4$^-$ with one stuck D4-brane.
We recall that its M-theory lift, exploited in this work, is $M_2\times (\IR^2\times \IX_6\times \IS^1)/\IZ_2$,
with the $\IZ_2$ including a half-period shift of the $\IS^1$.
Additional D4-brane pairs correspond to additional M5-brane pairs, as studied in the previous section.

\item An O4$^+$-plane (carrying $+1$ units of D4-brane charge).
Its lift to M-theory is a geometric orientifold $M_2\times (\IR^2\times \IX_6)/\IZ_2\times \IS^1$ with two stuck M5-branes on top.
The M5-branes are stuck because they do not form an orientifold pair, due to a different worldvolume Wilson line \cite{Gimon:1998be}.

\item An ${\widetilde{\rm O4}}^+$-plane, which can be regarded as an O4$^+$ with an extra RR background field.
Its M-theory lift is our $M_2\times (\IR^2\times \IX_6\times \IS^1)/\IZ_2$ geometry, with one stuck M5-brane fixed by the $\IZ_2$ action.
\end{itemize}

\subsubsection{The \texorpdfstring{O4$^-$ vs.\ the O4$^0$}{O4- vs. the O40} case}

Several references, e.g.\ \cite{Sinha:2000ap,Acharya:2002ag,Bouchard:2004ri,Bouchard:2004iu,Aganagic:2012au}, consider unoriented A-models with no open string sector, corresponding to an M-theory lift $(M_4\times \IX_6)/\IZ_2\times \IS^1$. This corresponds to the case of the O4$^-$-plane (for their relation with the O4$^+$ case, see later).
The key difference with our setup is that in general there are $\IZ_2$ fixed points,
which correspond to $L\times \IR^2\times \IS^1$.
The physics near these singularities cannot be addressed with present technology.
However, since crosscap embeddings do not intersect $L$,
it is possible to meaningfully propose an M-theory Gopakumar-Vafa interpretation of the unoriented topological string amplitude.
This can also be extended to open string sectors, as long as the D4-branes (or M5-branes in the M-theory lift)
are introduced in pairs and kept away from the singular locus.
As we discuss later on, this limits the possibility of reproducing the right physics for the O4$^+$.

A second difference from our setup is that the orientifold action in M-theory is felt even locally on the $\IS^1$,
i.e.\ already at the level of the 5d theory, and breaks the $\SU(2)_\L$ symmetry.
Therefore the structure of multiplets need not correspond to full $\SU(2)_\L$ multiplets, although this is explicitly assumed in most of these references.
Although supported by the appropriate integrality properties derived from the analysis,
these extra assumptions obscure the physical derivation of the BPS integrality structures.

We emphasize again that these properties differ in our system, which corresponds to the lift of the O4$^0$-plane.
The $\SU(2)_\L$ multiplet structure is directly inherited from the parent theory, and is therefore manifestly present, without extra assumptions.

\subsubsection{The \texorpdfstring{O4$^+$}{O4+} case}

The case of the O4$^+$-plane has been discussed in the literature as a minor modification of the O4$^-$ case.
Indeed, from the viewpoint of the type IIA theory (equiv.\ of the topological A-model),
both systems are related by a weight $(-1)^c$ for any worldsheet amplitude with $c$ crosscaps (corresponding to a change in the NSNS 2-form background around the orientifold plane). This motivates an immediate BPS invariant expansion of the unoriented A-model amplitude corresponding to the O4$^+$, see e.g.\ \cite{Sinha:2000ap,Acharya:2002ag,Aganagic:2012au}. 

On the other hand, the actual M-theory lift  of the O4$^+$-plane corresponds not to the geometry
$M_2\times (\IR^2\times \IX_6)/\IZ_2 \times \IS^1$
with a different choice of 3-form background,
but rather to the same geometry as the O4$^-$-plane, with the addition of two stuck M5-branes at the $\IZ_2$ fixed point.
In this lift, interestingly, the M-theory picture contains both unoriented and open M2-brane curves, which should combine together to reproduce a purely unoriented Gromov-Witten worldsheet expansion; the latter moreover admits a BPS expansion in terms of purely closed M2-brane curves, up to some sign flips. It is non-trivial to verify how these pictures fit together, in particular given the difficulties in dealing with M2-branes ending on M5-branes stuck at the $\IZ_2$ fixed point in the M-theory geometry. The details of this connection are therefore still open, and we leave them for future work.

\subsubsection{The \texorpdfstring{${\widetilde{\rm O4}}^+$}{tildeO4+} case}

Finally, the case of the ${\widetilde{\rm O4}}^+$-plane has not been considered in the literature.
Actually, it is closely related to the lift of the O4$^0$, with the addition of one M5-brane.
It is therefore very similar to the systems in the previous section,
and the corresponding amplitude is essentially given by \cref{final-trace-open},
for $2N+1$ M5-branes (allowing for the addition of $N$ brane pairs).

In this case there is also an interesting interplay with the type IIA picture,
although in the opposite direction as compared with the O4$^0$ case.
Namely, the M-theory lift contains one more brane than the corresponding type IIA picture.
This implies that in the BPS expansion both closed and open M2-brane states have to combine together
to reproduce the crosscap worldsheet diagram in type IIA.
It would be interesting to carry out this comparison further,
although this may be difficult due to the presence of a non-trivial RR background in the type IIA orientifold, which may render the worldsheet analysis difficult.

\section{Conclusions and open issues}
\label{sec:conclu}

In this paper we have discussed the BPS integer expansion of the real topological string in \cite{Walcher:2007qp}, using the M-theory lift of the O4-plane with one stuck D4-brane. Since the geometry is a $\IZ_2$ quotient acting freely in the M-theory $\IS^1$, the 5d setup enjoys an enhancement to 8 supercharges and is identical to that in the closed oriented Gopakumar-Vafa system. The subtleties due to the orientifold quotient arise as a compactification effect modifying the KK momentum of the BPS states on the $\IS^1$ according to their parity under the orientifold action. This allows for a clean derivation of the BPS integer expansion, without the extra assumptions that pop up in other unoriented A-models.

Although we recover the  BPS expansion conjectured in \cite{Walcher:2007qp}, our derivation shows the correct identification of the BPS invariants not as the equivariant sector of the parent Gopakumar-Vafa invariants, but rather a weighted version thereof. 

The M-theory picture provides a complementary viewpoint on the sign choices implied by the tadpole cancellations in models where the fixed lagrangian 3-cycle $L$ has $H_1(L;\IZ)=\IZ_2$ \cite{Walcher:2007qp}. More in general, the BPS integer expansion we propose is valid  for other situations, providing a general definition of the real topological string. 

The careful M-theory lift of other O4-planes suggests non-trivial relations between their BPS invariant expansions, for instance the addition of an open M2-brane sector (associated to two stuck M5-branes) to the lift of the O4$^-$-plane should reproduce a sign flip in odd crosscap contributions. This seems to imply non-trivial relations among the unoriented and open BPS invariants in M-theory orientifolds with fixed points. We hope to return to these and other questions in the future.

\subsection*{Acknowledgements}

We thank G. Bonelli, R. Gopakumar, D. Krefl, A. Tanzini, and J. Walcher for useful discussions.
AU is partially supported by the grants FPA2012-32828 from the MINECO,
the ERC Advanced Grant SPLE under contract ERC-2012-ADG-20120216-320421
and the grant SEV-2012-0249 of the ``Centro de Excelencia Severo Ochoa" Programme.
NP is partially supported by the
COST Action MP1210 ``The string Theory Universe''
under STSM 15772.

\appendix

\section{Review of Walcher's real topological string}
\label{sec:walcher}

We now give a short review of A-model localization as in \cite{Walcher:2007qp}. Take for concreteness Fermat quintic, given by
\beq \left\{ \sum_{i=1}^5 x_i^5 = 0 \right\} \subset \CP^4 \eeq
and involution
\beq \sigma: \quad (x_1:x_2:x_3:x_4:x_5) \mapsto (\ov{x}_2:\ov{x}_1:\ov{x}_4:\ov{x}_3:\ov{x}_5)\eeq
which gives a fixed point lagrangian locus $L$ with $\RP^3$ topology, hence $H_1(L;\IZ)=\IZ_2$.

\subsection{Tadpole cancellation in the topological string}

Requiring a function $f$ to be equivariant implies that
fixed points of $\Omega$ are mapped to $L$,
but one has to further specify their homology class in $H_1(L;\IZ)=\IZ_2$.
When that class is trivial, then under deformation of the map
it can happen that the boundary is collapsed to a point on $L$.\footnote
{This is a real codimension one stratum in the moduli space, as discussed in \cref{unoriented-gv}.}

The local model for this phenomenon is a Veronese-like embedding $\CP^1 \to  \CP^2$
defined by a map $ (u:v) \mapsto (x:y:z)$
depending on a target space parameter $a$, concretely
\beq x= au^2, \quad y=av^2, \quad z=uv. \label{veronese}\eeq
The image can be described as the conic $xy-a^2z^2=0$, and it is invariant under $\sigma$ if $a^2 \in \IR$.
The singular conic $a=0$ admits two different equivariant smoothings, determined by the nature of $a$:
\beq
\begin{aligned}[3] &a \in \IR \quad &(u:v)& \sim (\overline{v}:\overline{u}) \quad &\text{disk} \\
&a \in \iu \IR \quad &(u:v)& \sim (\overline{v}:-\overline{u}) \quad &\text{crosscap}.\end{aligned}
\eeq
The proposal of \cite{Walcher:2007qp} to account for this process
is to count disks with collapsible boundaries and crosscaps together.
Specifically, there is a one-to-one correspondence between even degree maps leading to boundaries and maps leading to crosscaps
(which must be of even degree, in order to be compatible with the antiholomorphic involution, as already manifest in the above local example).
The tadpole cancellation condition amounts to proposing the combination of these paired diagrams, such that certain cancellations occur.
For instance, the amplitude $\mathcal{R}^{(g,h)}$ (odd number of crosscaps) vanish
(due to the cancellation of the unpaired crosscap with an odd degree boundary).
Similarly, for the remaining contributions, in terms of $\chi$ and $d$, the tadpole cancellation imposes the restriction
\beq d \equiv h \equiv \chi \mod{2}.\eeq
The first equality follows from the requirement that
odd degree contributions only come from boundaries
(homologically trivial, i.e.\ even degree, ones cancel against crosscaps),
while the second from the requirement that there be no unpaired crosscaps.

\subsection{Rules of computation}

One can then \emph{postulate} the existence of a well-defined fundamental class
allowing to integrate over the moduli space $\ms = \ov{\ms}_{\Sigma} (\CP^4,d)$ of maps, defined as
the top Chern class of an appropriate bundle $\mathcal{E}_d$ over $\ms$,
whose fiber is given by $H^0(\Sigma,f^*\mathcal{O}(5))$:
\beq 
\tgw^\Sigma_d = \int_\ms \mathbf{e}(\mathcal{E}_d).
\eeq
For the integral to make sense, dimensions are constrained as
\beq
5d+1 = \dim \mathcal{E}_d \stackrel{!}{=} \operatorname{vdim} \overline{\mathcal{M} }_{g,n}(\CP^D,\beta) = c_1 \cdot \beta + (3-D)(g-1) +n,
\eeq
where $\beta \in H_2 (\CP^D;\IZ)$ is to be identified with $d$ in this particular case,
and $n$ denotes the number of punctures.

The next step is to apply Atiyah-Bott localization to the subtorus $\IT^2 \subset \IT^5$ compatible with $\sigma$.
As explained in \cite{Kontsevich:1994na}, the fixed loci of the torus action are given by nodal curves,
in which any node or any component of non-zero genus is collapsed to one of the fixed points in target space,
and any non-contracted rational component is mapped on one of the coordinate lines with a standard map
of given degree $d_i$:
\beq
f(w_1:w_2) = (0:\ldots:0:w_1^{d_i}:0:\ldots:0:w_2^{d_i}:0:\ldots:0). \label{ratmap}
\eeq
The components of the fixed locus can be represented by a decorated graph $\Gamma$
and one has well-defined rules for associating a graph to a class of stable maps.

For the case of real maps,
one has to be extra careful and require the decoration to be compatible with the action of $\Omega$ and $\sigma$.
For example, consider a fixed edge:
if we think of $z=w_1/w_2$,
then in \cref{ratmap} $z \mapsto 1/\bz$
is compatible (i.e.\ $f$ is equivariant) with any degree, while
$z \mapsto -1/\bz$
requires even degree,
because our involution acts on the target space
as $(x_1:x_2:\ldots)\mapsto (\overline{x}_2:\overline{x}_1:\ldots)$
i.e.\ $w=x_1/x_2 \mapsto 1/\overline{w}.$

\bigskip

A careful analysis
in \cite{Walcher:2007qp,Krefl:2009md,Walcher:2006rs,pandharipande-solomon-walcher}
allows to conclude that
the localization formula takes the form
\beq \label{orientability} \tgw^\Sigma_d = (-1)^{p(\Sigma)} \sum_\Gamma \frac1{\operatorname{Aut}\Gamma} \int_{\ms_\Gamma}
\frac{\mathbf{e}(\mathcal{E}_d)}{\mathbf{e}(\mathcal{N}_\Gamma)},\eeq
where the following prescriptions are used:
\begin{enumerate}
\item the $(-1)^{p(\Sigma)}$ factor in front of the localization formula
is put by hand in order to fix the relative orientation between different components of moduli space;

\item for any fixed edge of even degree, the homologically trivial disk and crosscap contributions
are summed, with a relative sign such that they cancel.
This is the above mentioned \emph{tadpole cancellation}.

\end{enumerate}

Finally, \cite{Walcher:2007qp} proposes an integer BPS interpretation of the obtained rational numbers $\tgw$:
by combining them with the above prescribed signs at fixed $\chi$, one gets integer numbers
$\tgv$, which are conjectured to reproduce a BPS expansion for the open-unoriented topological amplitudes.

\section{Physical couplings}
\label{sec:couplings}
An interesting question one can ask is what kind of coupling the topological string is computing
in the IIA physical theory.
This has a clear answer for the closed oriented sector \cite{Antoniadis:1993ze,Bershadsky:1993cx},
while some proposals have been made for open oriented \cite{Ooguri:1999bv}
and unoriented sectors\cite{Sinha:2000ap}.
Here we make a proposal for the analogous expression in our unoriented model.

We have 4 supercharges in $1+1$ dimensions,
and we'd like to find a good splitting of the $\NN=2$ Weyl tensor
\beq \mathcal W^{ij}_{\mu\nu} = T^{ij}_{\mu\nu} + R_{\mu\nu\rho\lambda} \theta^i \sigma^{\rho\lambda} \theta^j + \cdots, \eeq
where one requires the graviphoton field strength $T$
to acquire a self-dual background,
and hence can write
\beq \mathcal W_{\a\b} =
\frac12 \vep_{ij} \mathcal W^{ij}_{\mu\nu} (\sigma^\mu)_{\a\ad} (\sigma^\nu)_{\b\bd} \vep^{\ad\bd}.\eeq

A natural guess is that the amplitude $\mathcal G^{(\chi)}$ as in \cref{total-g} computes
\beq
\int \ud^4 x \int \ud^4 \theta \, \delta^2(\theta) \delta^2(x)
\left( \mathcal G^{(\chi)}(t) - \frac12 \mathcal{F}^{(g_\chi)}(t) \right) ( \mathcal W \cdot v)^{\hat g},
\eeq
where
$ \mathcal W\cdot v= \mathcal W_{\a\b} \sigma^{\mu\nu}_{\a\b} v_{\mu\nu}$.
This has a contribution $RT^{\hat g-1}$
(in the covering picture $\hat g-1=\chi$)
which in principle can generate (via SUSY)
the $\sinh^{-1}$ power in the Schwinger computation,
taking into account the fact that
the orientifold halves the number of fermion zero modes
on the Riemann surface.

It would be interesting to discuss the appearance of this contribution from different topologies at fixed $\chi$.
We leave this for future work.


\bibliographystyle{JHEP}
\bibliography{biblio}
\end{document}